COPYRIGHT NOTICE

# Electro-acoustic Scattering from a Pulsating Sphere

V. Viswarupa, Yoginder Kumar Negi, *Member, IEEE,* and N. Balakrishnan

*Abstract*—In this paper, we show the RCS enhancement due to the acoustic disturbances around a pulsating sphere. The acoustic variation is modelled with the dielectric inhomogeneities around the sphere caused by the pressure fluctuations due to the acoustic source. RCS is computed for the modelled dielectric pulsating sphere, a cube and a cone on a cylinder across a frequency band using Finite Difference Time Domain (FDTD) method. The RCS of the pulsating sphere and other objects considered are dominated by the background scattering from the pulsating object. In this work, we show that the dielectric variation due to the acoustic source can be detected even if there is no scattering from the object . The scattering from the dielectric variation leads to the detection of Bragg scattering along with a significant increase in RCS.

*Index Terms*— Finite Difference Time Domain Method, FDTD, Radar Cross Section, RCS, acoustic, electro-acoustic scattering

## I. INTRODUCTION

Often the designers of military aircraft strive hard to reduce the Radar Cross Section (RCS) to avoid detection by enemy radars to increase survivability. Some of the RCS reduction techniques include shaping [1,2], Radar Absorbing Materials (RAM) [3], passive [4,5], and active cancellation [6] of scattered fields. Targets can be shaped in a manner to deflect scattered energy away from the radar. Although shaping is one of the most often applied RCS reduction techniques, there are constraints due to specific mission objectives that need to be met. The purpose of RAM, as the name suggests, is to absorb the incident energy and thereby reduce the amount of energy scattered in the direction of the radar. Passive cancellation refers to the addition of scatterers to cancel reflection from the object, also called impedance loading. The scattered field is a superposition of the fields scattered by the added scatterers and the object and is controlled and reduced by passive cancellation. In active cancellation of the scattered field, the target emits a signal at the same time as the one to be received by the radar with a suitable amplitude and out of phase with the radar signal. This results in the radar, in an ideal case receiving no signal, thus rendering the target invisible.

The advancement in stealth technology and the growing success of RCS reduction techniques have further encouraged the development of RCS enhancement techniques as countermeasures. Enhancement of RCS despite cloaking and low visibility is most sought after. [7,8,9] present some of the RCS enhancement techniques. Active enhancement techniques are especially useful in surveillance and road safety applications [8]. Passive enhancement techniques are challenging as an enhancement in RCS is required even in the absence of a scatterer [9]. All the above techniques may not be suitable for non-cooperative enemy targets.

Motivated by the success of using Radio Acoustic Sounding Systems (RASS) in detecting echoes from clear air, radar engineers have been focusing on detecting echoes from the medium surrounding the aircraft caused by fluctuations from the acoustic disturbances caused by aircraft wake and engine noise. The traditional approach is to use a frequency scanning radar that maximizes return using Bragg Scatter [10,11]. Ref. [10] presents an analytical method to analyze the interaction between electromagnetic and acoustic waves. However, the analytical approach is limited as it is applicable to canonical and simple geometric problems. In this paper, we show the use of high-resolution pulsed radar to demonstrate that echoes from acoustic disturbances indeed increase detectability. Since FDTD is the most natural choice for analyzing the narrow pulse incidence, this has been used in the analysis. The RCS results for acoustic source-enabled sphere, cube, and model missile depicted in a simple form of a cone on a cylinder show very clearly the enhancement from the acoustic disturbances is indeed significant.

The paper is organized as follows: Section II gives the details of the numerical approach followed; Section III elaborates on the electro-acoustic modelling for the electro-acoustic scattering computation. Numerical results and conclusions are discussed in Sections IV and V.

## II. NUMERICAL APPROACH

An acoustic source creates a pressure variation in its surrounding media which leads to a spherical sinusoidal dielectric inhomogeneity radially around the source. Electro-acoustic scattering is the scattering of EM waves due to these inhomogeneities in the medium. Acoustic pressure variations have a direct impact on the dielectric inhomogeneities in the surrounding medium. This phenomenon is used to observe wind speed and direction using a method proposed in [12,13].

Other inspired research topics are atmospheric profiling, wave propagation in turbulent media, and Radio Acoustic Sounding Systems (RASS). RASS systems combine acoustic and radar techniques to measure the speed of propagation of an acoustic disturbance. This technique has been used in the observation of wind velocity and direction and indoor and outdoor temperature profiling [14,15].

In a RASS, the acoustic source is placed near the radar to

This work was supported by the Department of Science and Technology (DST) Government of India under National Supercomputer Mission (NSM) project SP/DSTO-20-0130,

V. Viswarupa, Yoginder Kumar Negi and N. Balakrishnan are with the Supercomputer Education Research Centre (SERC) at the Indian Institute of Science, Bengaluru, India. (Email: viswa.rupa98@gmail.com, yknegi@gmail.com, balki@iisc.ac.in)

produce the inhomogeneities, whereas, in our work, we consider the target itself to be an acoustic source and determine the changes in RCS due to acoustic disturbances it generates. Such an electro-acoustic scattering requires the solution for scattered fields from a target subject to a plane electromagnetic wave and can be harnessed to enhance target RCS. The paper [16]shows the suggested enhancement in the RCS by Bragg scattering for a pulsating metallic sphere, and their analysis is based on Method of Moments (MoM) [17] solutions which are computationally expensive, as discussed in the following.

Maxwell's equations are the most elegant expressions to solve such an electromagnetic problem. Solving Maxwell's equations analytically is hard due to boundary and interface conditions. Hence, numerical methods are employed to solve them. Numerically, Maxwell's equations can be solved either in the frequency domain or the time domain. Furthermore, these can be applied to the Partial Difference Equations (PDE) or the Integral Equation (IE) formulations of Maxwell's equations. Frequency-domain formulations such as Finite Element Method (FEM) [18] and MoM [17] are employed to solve electromagnetic problems from the point of view of frequency, often giving an intuitive understanding of the behavior of the system. However, solutions to Maxwell's equations in the frequency domain become increasingly complicated with the introduction of inhomogeneities, non-linearities, and complex geometries. The time domain formulations such as Finite Difference Time Domain (FDTD) [19], Finite Volume Time Domain (FVTD) [20], Finite Element Time Domain (FETD) [21], Plane Wave Time Domain (PWTD) [22] methods overcome the challenges posed by the frequency domain methods. They can solve a problem for several frequencies in a single calculation and are most natural for solving electromagnetic scattering problems for pulse incidence.

FDTD [23,24]is the most used time domain method due to its simplicity and unsophistication. FDTD is a versatile method for problems involving frequency responses over long ranges and overcomes the instability issues present in time-domain integral equations. The computational domain of FDTD consists of a custom geometry subject to excitation. The FDTD simulation of an open problem, as required in the case at hand, is achieved by the introduction of reflectionless absorbing boundaries [25,26,27,28]. We use the open-source FDTD solver OpenEMS [29] for the inhomogeneous body scattering analysis. To mimic real-life transient sources, plane wave excitation is introduced to the simulation domain using Total-Field Scattered-Field (TFSF) [30] boundary formulation. OpenEMS is written in C++ software with a MATLAB interface. Figure 1 shows the bi-static RCS of a sphere computed using OpenEMS and the Mie Series solution for a dielectric sphere of radius 0.1m and dielectric constant ($\epsilon_r$) of 1.5. The computed RCS shows that the accuracy of the OpenEMS is very good.

In this paper, the work is built upon our previous paper [31] and extends our investigation to higher frequencies and a greater number of unknowns, and delves into the exploration of Bragg scattering, as detailed in the next section.

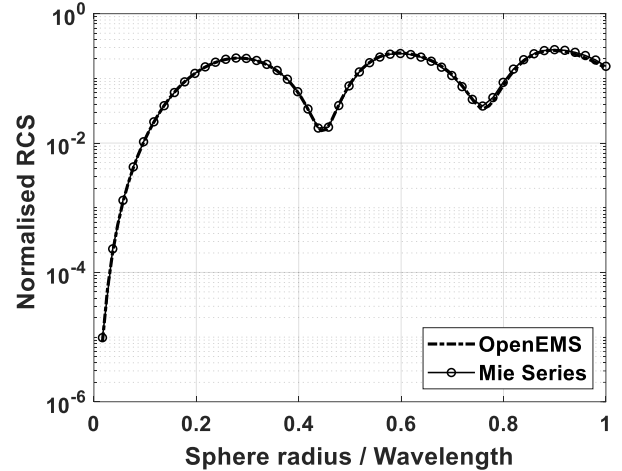

**Fig. 1.** RCS plots obtained from OpenEMS and Mie Series.

### III. ELECTRO-ACOUSTIC MODELLING

Sound is a disturbance propagating through an elastic medium leading to changes in pressure or displacement of the particles of the medium, which can be detected by an instrument [32]. The three main deductions from the description of sound by the Navier Stokes equations are that (I) sound at time $t$ depends on sound at $t - r/c$.(where r is the distance from the acoustic source and c is the speed of sound in the medium) (II) shape of the wave remains unchanged. (III) the magnitude of pressure varies (decreases) as the wave propagates. The simplest acoustic model that can be considered is the sound radiated by a pulsating sphere. Considering a sphere of radius $a$ to vibrate at a frequency $\omega$ with an amplitude $V$, its surface radial velocity $v$ and pressure $p$ can be written as,

$$v = Ve^{-j\omega t} \quad (3.1)$$

$$p = Pe^{-j\omega t} \quad (3.2)$$

Applying equations (3.1) and (3.2) to the linearized acceleration-pressure gradient relationship, the solution can be obtained as,

$$p = f(t - r/c)/r = Ae^{-j\omega(t-r/c)}/r \quad (3.3)$$

Where $r$ is the distance from the sphere, $t$ is time. The solution for pressure is then [33],

$$p = \frac{k_a a}{r} \frac{k_a a - j}{(k_a a)^2 + 1} (\rho_0 V c_a a) e^{-jk(r-a)} e^{-j\omega t} \quad (3.4)$$

where $\rho_0$ is the equilibrium density of the medium, $c_a$ is the velocity of sound in the medium, $k_a$ is the acoustic wavenumber, $\omega_a$ is the acoustic frequency, $r$ is the distance from the center of the pulsating sphere, and the parameter $k_a a$ is a non-dimensional characteristic called compactness. $k_a a < 1$ when the sphere is small or the frequency is low.

$k_a a > 1$ when the sphere is large or is vibrating at a high frequency. The velocity amplitude $V$ is given by,

$$V = \frac{P}{\rho_0 c_a} \quad (3.5)$$

where $P$ is related to the Sound Pressure Level (SPL) by,

$$SPL = 20 \, log \, log \, \frac{P}{P_0} \quad (3.6)$$

where $P_0$ is the reference pressure level (20 µPa is the reference sound pressure of air). Equations (3.6) and (3.5) show that the loudness of the sound affects the velocity amplitude of the acoustic wave. The velocity amplitude further affects the pressure variation around the acoustic source, as shown in equation (3.4). As can be observed in equation (3.4), the pressure variation is sinusoidal in nature. These sinusoidal alterations cause the $\varepsilon_r$ of the medium to vary sinusoidally in the absence of water vapor as [34],

$$\varepsilon_r = 1 + 1.5512 \times 10^{-6} \frac{p}{T} \quad (3.7)$$

Where $p, T$ are the real part of pressure in Pa and temperature in K, respectively.

RCS is maximized when the incident EM wave and the acoustic wave interfere constructively. According to Bragg's condition for the mono-static case, the wavelength of the incident EM wave must be twice the wavelength of the acoustic wave for the maximum scattered field to be observed. Mathematically this can be expressed as [35]

$$\lambda_e = 2\lambda_a \quad (3.8)$$

where, $\lambda_a$ and $\lambda_e$ are acoustic and incident EM wavelengths, respectively. When this condition is met, the acoustic and EM waves interfere constructively with each other, resulting in a maximum scattered field and hence a maximum RCS. Due to the different time scales of acoustic and EM waves, coupled acoustic and EM equation is difficult to model and solve. Hence, the electro-acoustic model consists of a pulsating spherical core frozen at time $t$, resulting in a spherical core with dielectric layers, as discussed in the following section.

## IV. NUMERICAL RESULTS

The acoustic frequency is selected for the highest maximum dielectric variation and the lowest minimum of $\varepsilon_r$. Louder acoustic disturbance gives larger pressure variation and enhances the detectability of the inhomogeneous media. The loudness of the acoustic signal or the Sound Pressure Level (SPL), measured in dB, is suitably selected. The variations due to the first wavelength alone are considered due to system constraints. In this paper, a 2m spherical acoustic source is considered. Electro-acoustic modelling is done for an acoustic frequency of 1000Hz and SPL= 194dB, resulting in a layered sphere of the outermost radius of 2.343 m, with each layer having a different dielectric constant. These dielectric constants are obtained from pressure values calculated using equations (3.4, 3.5, 3.6) substituted in equation (3.7). The layered spherical geometry, as shown in Figures 3, 4, 5 is exposed to a plane EM Gaussian pulse. To ensure that the enhancement in RCS is particularly due to the presence of an acoustic disturbance, the RCS plot from the electro-acoustic model is compared to that obtained from the same model without the acoustic disturbance. The geometry is modeled using FreeCAD [36] and then imported for simulation to the OpenEMS-MATLAB interface for electro-acoustic scattering computation. The Bragg frequency calculated using equation (3.8) is 437.32MHz. To verify the Bragg peak, a plot of the differences in RCS values with and without the acoustic disturbances is plotted and subjected to a spline fit.

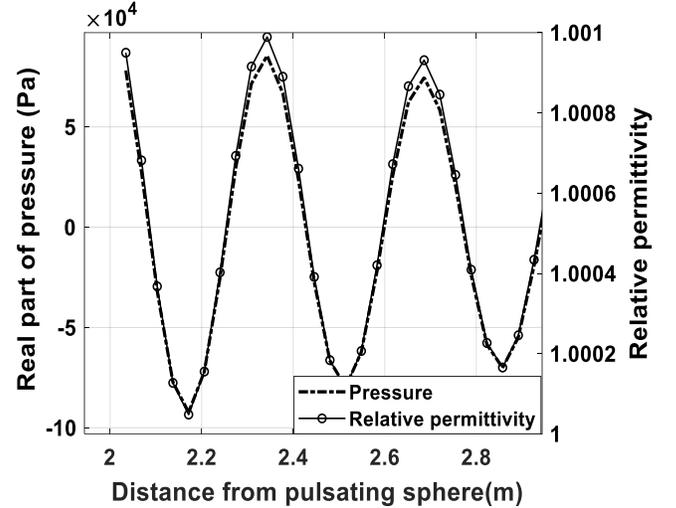

**Fig. 2.** Pressure and dielectric variation from 2m sphere for acoustic frequency of 1000Hz and SPL=194dB

*A. Spherical acoustic source*

Considering only the first acoustic wavelength, the dielectric layers span upto $1\lambda_a$ external to the spherical acoustic source of 2m. The geometry shown in Figures 3 and 4, 5 is subjected to a Gaussian pulse of 1.54ns duration with 33,076,161 FDTD cells. The RCS is computed over a range of EM frequencies between 200MHz - 600MHz while keeping the acoustic frequency constant at 1000Hz in order to observe Bragg's scattering. Figures 6, 7, 8 and 9 show the RCS plots for 2.0 m core of relative permittivities, 1.02, 1.01, 1.005, and 1.000, respectively. The difference in RCS levels with and without the presence of an acoustic source is plotted and analyzed, as shown in Figure 10. It can be observed from Figures 6,7,8,9 that as the core dielectric constant decreases, the RCS in the presence of acoustic disturbances increases significantly. On calculating the difference in RCS (dB) at each frequency and computing the average, a 37.364dB increase in RCS is found for the core. $\epsilon_r = 1.00$. For the $\epsilon_r$ values 1.02, 1.01, and 1.005, the increase in RCS is less pronounced, as evidenced by the greater overlap of the plots due to greater background scattering from the core.

The simulation to realize Bragg's scattering principle revealed a shift of about 60MHz from the expected Bragg frequency. This is due to the fact that the Bragg frequency is calculated using wavelengths in free space, and in actuality, the waves are traveling in a medium with dielectric constants modulated by acoustic waves.

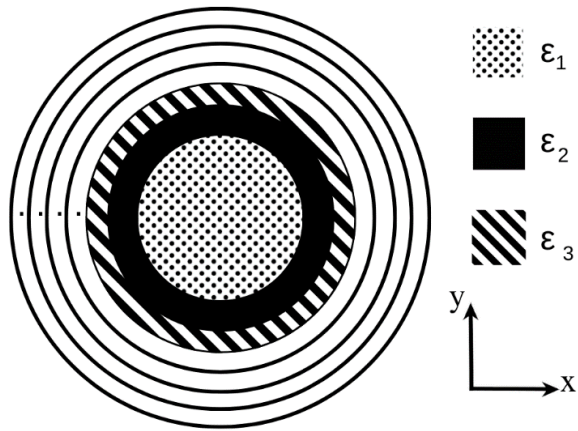

**Fig. 3.** 2D depiction of electro-acoustic model of a spherical acoustic source. The initial three dielectric layers have been visually depicted and the subsequent layers are allotted dielectric values similarly.

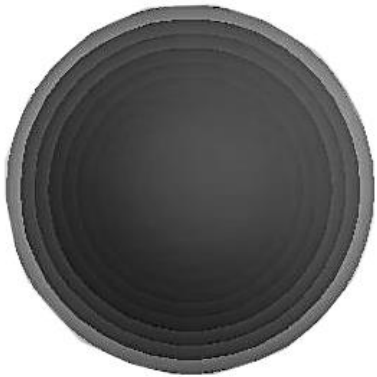

**Fig. 4.** 3D depiction of the pulsating core and the resulting dielectric layers around it.

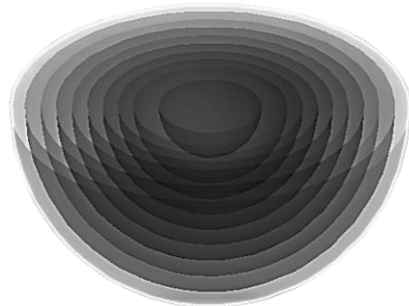

**Fig. 5.** Cross section of the pulsating sphere and the dielectric layers around it.

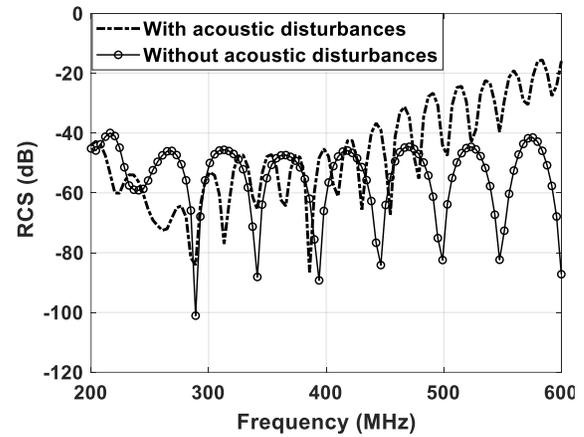

**Fig. 6.** Mono-static RCS of a layered sphere with and without acoustic disturbance for core dielectric ($\epsilon_r$=1.02).

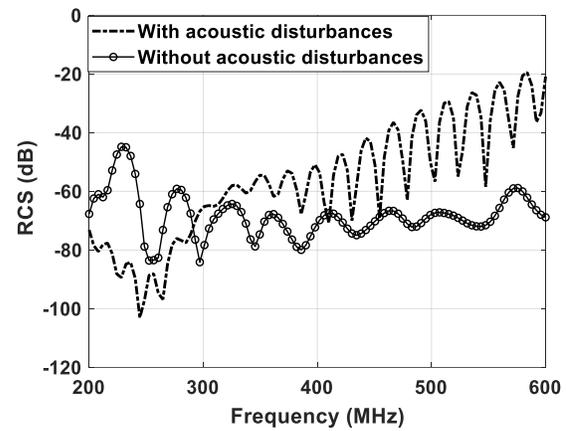

**Fig. 7.** Mono-static RCS of a layered sphere with and without acoustic disturbance for core dielectric ($\epsilon_r = 1.01$).

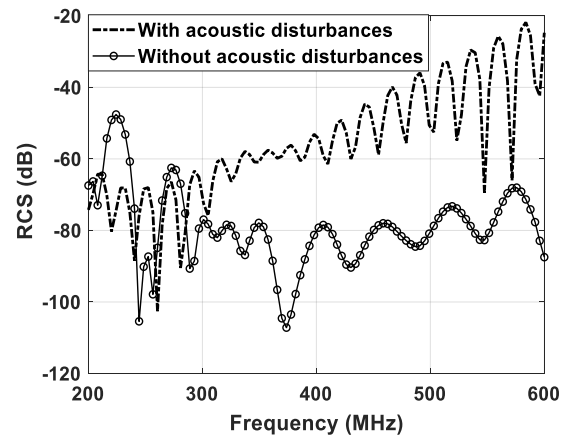

**Fig. 8.** Mono-static RCS of a layered sphere with and without acoustic disturbance for core dielectric ($\epsilon_r = 1.005$).

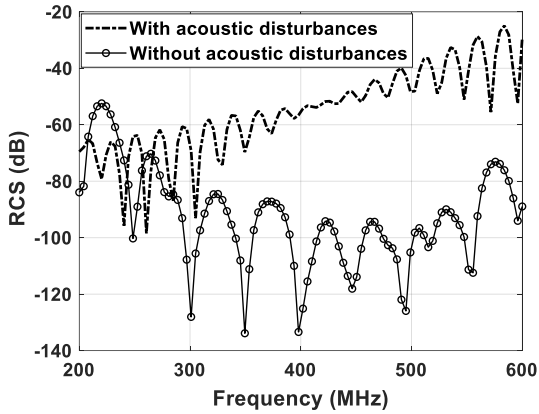

**Fig. 9.** Mono-static RCS of a layered sphere with and without acoustic disturbance for core dielectric ($\epsilon_r = 1.00$).

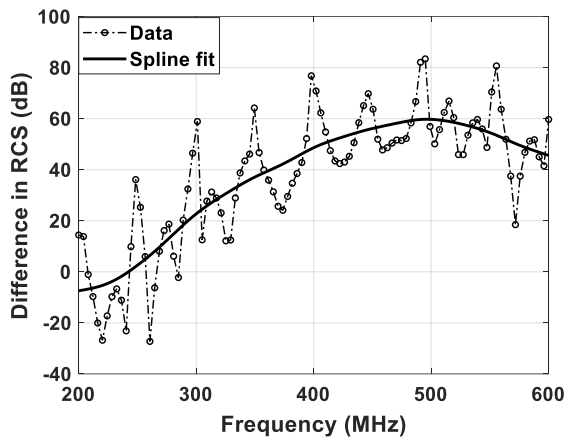

**Fig. 10.** RCS difference and spline fit with and without acoustic disturbance for core dielectric ($\epsilon_r = 1.00$).

It can be observed that Bragg's peak is visible for the core of $\varepsilon_r = 1.000$, in the further subsection, RCS is computed for the cubic and missile geometries for $\varepsilon_r = 1.000$.

### B. Cube

A cube of sides $(2 + \lambda_a)$m is considered such that it encloses the spherical acoustic source and the dielectric variations caused by the first acoustic wavelength, as shown in Figures 11 and 12. The geometry is subjected to a Gaussian pulse of 1.54ns duration.6,84,17,900 FDTD cells were considered. Figure 13 shows the RCS plots of a cube with and without the acoustic disturbance. The average enhancement calculated over the range of frequencies considered is 48.32 dB. Figure 14 is the plot of the differences in the RCS values with and without the acoustic disturbance (in dB) at each frequency. A spline fit of the data points reveals a Bragg's peak. The peak is observed to have shifted 60MHz from the expected frequency.

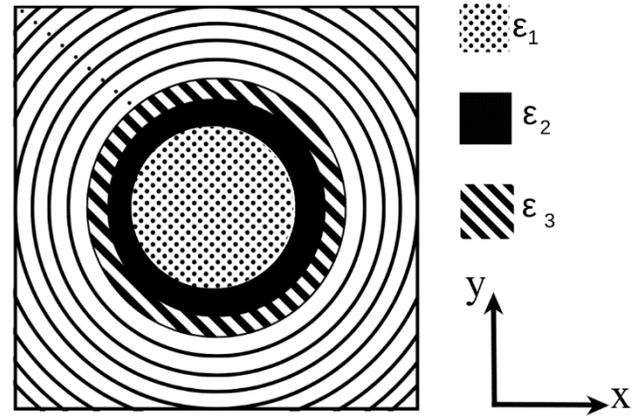

**Fig. 11.** 2D depiction of electro-acoustic model of the cube with a spherical acoustic source at its core and the dielectric layers around it. The initial three dielectric layers have been visually depicted and the subsequent layers are allotted dielectric values similarly.

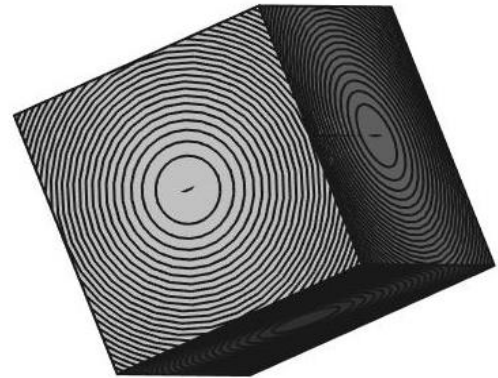

**Fig. 12.** 3D depiction of the cube with a spherical acoustic source at its core and the resulting dielectric layers around it

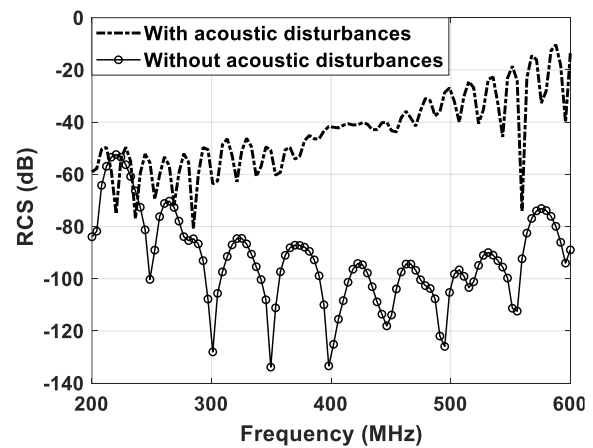

**Fig. 13.** Mono-static RCS of the cube of $\varepsilon_r = 1.000$ with a spherical acoustic source at its body center.

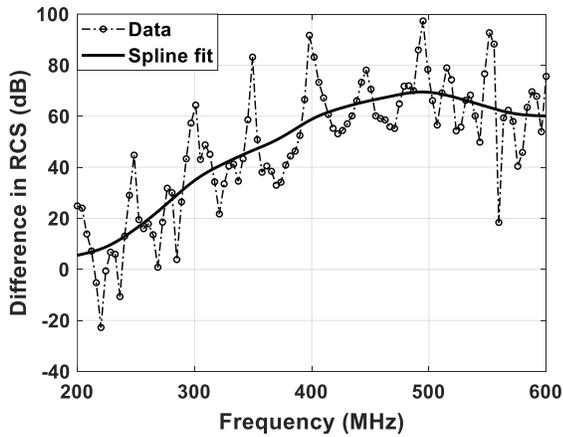

**Fig. 14.** RCS difference and spline fit with and without acoustic disturbance for cube dielectric ($\epsilon_r = 1.00$).

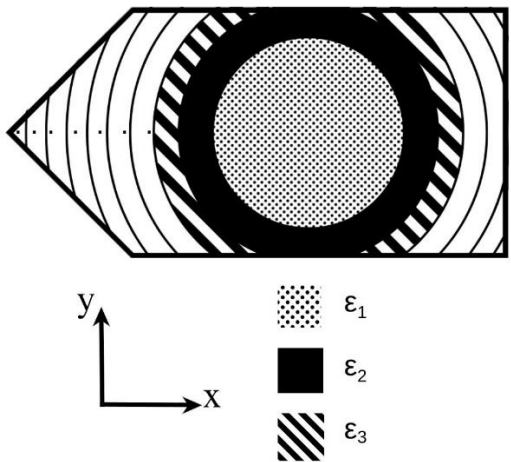

**Fig. 15.** 2D depiction of electro-acoustic model of the missile with a spherical acoustic source at its body center and the dielectric layers around it. The initial three dielectric layers have been visually depicted and the subsequent layers are allotted dielectric values similarly.

*C. Missile model*

The missile model has a radius of 2.343m, chosen to fully encompass both the acoustic source and the first acoustic wavelength. The length from the nose point to the base of the model measures 6.4 m. The model is subjected to a Gaussian pulse of 1.54ns. The incident plane wave direction is aligned with the y-axis, i.e., perpendicular to the body axis of the model. The number of FDTD cells considered is 4,86,27,100 in the case of the model with no acoustic disturbance and 16,72,84,000 for the model with an acoustic source. Figure 17 shows the RCS plots of the missile. Figure 18 is a plot of the differences in RCS values with and without acoustic sources. The average increase in RCS is observed to be 5.7 dB. In this case of the model missile, the Bragg shift is slightly lesser than compared to the cube and sphere. This is because of complex structure modeling issues due to system constraints. The theoretical EM Bragg frequency for the acoustic disturbance of 1000Hz is 437.32MHz. However, from the above figures, we see that it is 494MHz. This small difference has risen due to the following. (I) Theoretical Bragg scattering is calculated assuming free space. (II) Due to numerical limitation, the Bragg frequency from Figures 10, 14, and 18 is observed after smoothening the RCS curve and hence is much coarse.

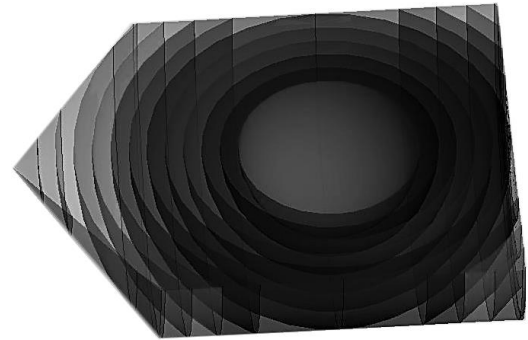

**Fig. 16**. 3D depiction of a missile model with a spherical acoustic source at its core and the resulting dielectric layers around it.

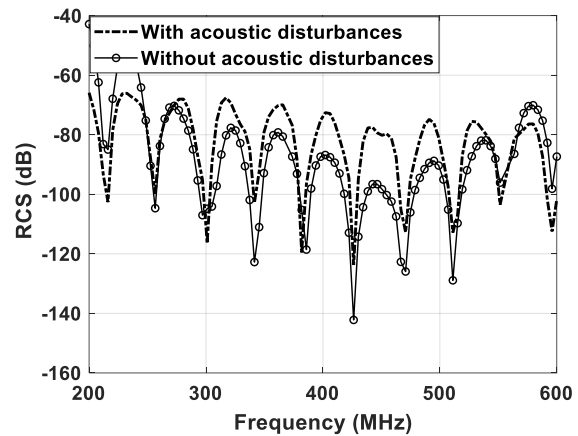

**Fig. 17.** Mono-static RCS of missile model of $\epsilon_r = 1.000$ with a spherical acoustic source at its body center.

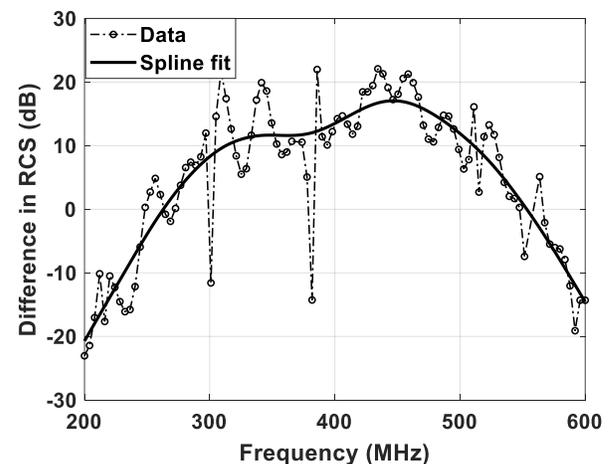

**Fig. 18.** RCS difference and spline fit with and without acoustic disturbance for cube dielectric ($\epsilon_r = 1.00$).

## V. Conclusions

In this paper, a numerical approach to an electro-acoustic system by modelling a pulsating sphere demonstrates that these acoustic disturbances can indeed enhance the Radar Cross Section (RCS). The results from computing RCS for a cubic geometry reveal the same despite the increased complexity of the geometry. Specifically, the dielectric variation in the surrounding medium due to the acoustic source can be detected efficiently in the absence of background scattering and thus aid in the detection of the seemingly invisible aircraft. Our results show the detection of Bragg scattering in the presence of acoustic disturbances with a slight shift in the Bragg peak. The same pulsating sphere has been used inside the cube and missile model; it is only logical that the ratio of the numerically computed Bragg frequency to the theoretical Bragg frequency turns out to be 1.13 in both cases. The shift from 1 is due to the consideration of free space in theoretical Bragg frequency, and spline approximation is the RCS plot. We suspect that another cause for the deviation from the expected Bragg frequency in the current work is due to considering a single wavelength electro-acoustic geometry, software, and system constraints. In future work, the RCS enhancement will be explored further by modelling a larger electro-acoustic system of multiple wavelengths.

Despite the current development of wide-band RAM materials, which render fighter aircraft invisible to radar, every aircraft has a unique acoustic signature at different atmospheric altitudes. The method proposed can be used to detect the various types of aircraft. In fact, the acoustic signatures are unique to the type of aircraft and hence the Bragg frequency. This technique, in effect, can enhance not only the detection of aircraft but also their identification.

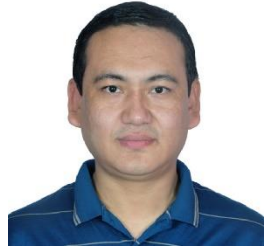
**Yoginder Kumar Negi** obtained the B.Tech degree in Electronics and Communic-ation Engineering from Guru Gobind Singh Indraprastha University, New Delhi, India, in 2005, M.Tech degree in Microwave Electronics from Delhi University, New Delhi, India, in 2007 and the PhD degree in engineering from Indian Institute of Science (IISc), Bangalore, India, in 2018.

Dr. Negi joined Supercomputer Education Research Center (SERC), IISc Bangalore in 2008 as a Scientific Officer. He is currently working as a Senior Scientific Officer in SERC IISc Bangalore. His current research interests include numerical electromagnetics, fast techniques for electromagnetic application, bio-electromagnetics, high-performance computing, and antenna design and analysis.
.

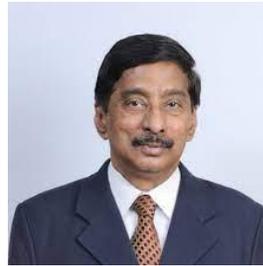
**B Narayanaswamy** received the B.E. degree (Hons.) in Electronics and Communi-cation from the University of Madras, Chennai, India, in 1972, and the Ph.D. degree from the Indian Institute of Science, Bengaluru, India, in 1979.

He joined the Department of Aerospace Engineering, Indian Institute of Science, as an Assistant Professor, in 1981, where he became a Full Professor in 1991, served as the Associate Director, from 2005 to 2014, and is currently an INSA Senior Scientist at the Supercomputer Education and Research Centre. He has authored over 200 publications in the international journals and international conferences. His current research interests include numerical electromagnetics, high-performance computing and networks, polarimetric radars and aerospace electronic systems, information security, and digital library.

Dr. Narayanaswamy is a fellow of the World Academy of Sciences (TWAS), the National Academy of Science, the Indian Academy of Sciences, the Indian National Academy of Engineering, the National Academy of Sciences, and the Institution of Electronics and Telecommunication Engineers.

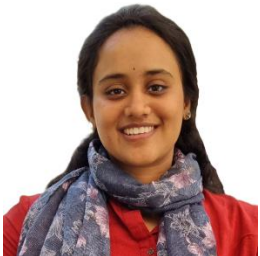
**Viswarupa. V** obtained the B.Sc. (Hons.) degree in Physics from Sri Sathya Sai Institute of Higher Learning, Anantapur, India, in 2019, M.Sc. degree in Electronics from Mount Carmel College, Bengaluru, India, in 2021. She joined Supercomputer Education Research Center (SERC), IISc Bengaluru in 2022 and is currently working as a Project Associate here. Her current research interests include electromagnetics, numerical electromagnetics and bio-electromagnetics.